\documentclass[11pt,twoside]{article}
\usepackage{newpasp,psfig}
\def\edcomment#1{\iffalse\marginpar{\raggedright\sl#1\/}\else\relax\fi}
\reversemarginpar

\def\solar{\ifmmode_{\mathord\odot}\else$_{\mathord\odot}$\fi~}
\def\hub{$H_0=100$ km s$^{-1}$ Mpc$^{-1}$, $q_0 = 0.5$}
\def\sprop{$S_{\nu} \propto \nu^{\alpha}$}

\def\deg{\ifmmode $\setbox0=\hbox{$^{\circ}$}$^{\,\circ}
          \else    \setbox0=\hbox{$^{\circ}$}$^{\,\circ}$\fi\,}

\begin{document}
\title{A High-Frequency and Multi-Epoch VLBI Study of 3C\,273}
\author{
T.P. Krichbaum$^1$,
D.A. Graham$^1$, 
%A. Lobanov$^1$,
A. Witzel$^1$,
J.A. Zensus$^1$,
A. Greve$^2$,
M. Grewing$^2$,
A. Marscher$^3$,
and A.J. Beasley$^4$
}
\affil{
(1) Max-Planck-Institut f\"ur Radioastronomie, Bonn, Germany\\
(2) Institut de Radioastronomie Millim\'etrique, Grenoble, France\\
(3) Department of Astronomy, Boston University, Boston, USA\\
(4) National Radio Astronomy Observatory, Socorro, USA\\
}

\begin{abstract}
We show results from a 7 year VLBI monitoring programme of 3C\,273 
at millimeter wavelengths.
We find evidence for component acceleration, motion or rotation of fluid 
dynamical patterns, and an outburst-ejection relation between 
$\gamma-$ray flares and new jet components.
\end{abstract}

\vspace{-1cm}
\section{Introduction}
\vspace{-0.3cm}
VLBI imaging at millimeter wavelengths (mm-VLBI) gives highest angular resolution
and facilitates direct studies of the self-absorbed jet base in AGN. For 3C\,273 ($z=0.158$)
images with an angular resolution of up to $50$\,$\mu$as (1\,$\mu$as $= 10^{-6}$\,arcsec) are made
at 86\,GHz. This corresponds to a spatial scale of $\sim 1000$ Schwarzschild radii
(for M $=10^9$ M\solar). 
The small observing beam allows to accurately determine positions of jet components.
This facilitates detailed studies of the jet structure
and kinematics near the nucleus, in particular with regard to the broad-band (radio to Gamma-ray)
flux density variability and the birth of new `VLBI components'. Here we summarize
new results from our multi epoch (1990 -- 1997) and high observing frequency 
(15, 22, 43 and 86\,GHz) VLBI monitoring.

\vspace{-0.5cm}
\section{Results}
\vspace{-0.3cm}
At sub-milliarcsecond resolution, 3C\,273 shows a one sided core-jet structure of several
milliarcseconds length. The jet breaks up into multiple VLBI components, which -- when
represented by Gaussian components -- seem to separate at apparent superluminal
speeds from the stationary assumed VLBI core. The cross-identification of the model-fit components,
seen at different times and epochs, is facilitated by small ($< 0.2$\,mas), and to first
order negligible, opacity shifts of the component positions relative to the VLBI-core. 
Quasi-simultaneous data sets (cf. Fig.\ 1) demonstrate convincingly the reliability of the component 
identification, which results in a kinematic scenario, in which all detected jet components
(C6 -- C18) move steadily (without `jumps' in position) away from the core (Fig.\ 4).
For the components
with enough data points at small ($<2$\,mas) and large ($> 2$\,mas) core separations, quadratic fits
to the radial motion r(t) (but also for x(t) --right ascension, and y(t) --declination) represent
the observations much better than linear fits.
Thus the components seem to accelerate as they move out. The velocities range typically from
$\beta_{app} =4 -8$ (for \hub).

\begin{figure}[p]
\begin{minipage}[t]{6.5cm}{
\psfig{figure=kri_fig1.epsi,width=5.7cm}
%\caption
{\footnotesize Fig.\ 1:
3C\,273 at 22\,GHz (top), 43\,GHz (center),
and 86\,GHz (bottom) observed in  1995.15 -- 1995.18.
Contour levels are -0.5, 0.5, 1, 2, 5, 10, 15, 30, 50, 70, and
90\,\% of the peak of 3.0 (top), 5.4 (center), and 4.7\,Jy/beam (bottom).
For the 22\,GHz map, the 0.5\,\% contour is omitted. The restoring beam is
$0.4 ~{\rm x}~ 0.15$\,mas in size, oriented at $\rm{pa}=0 \deg$. The maps are centered on the
eastern component (the core), the dashed lines guide the eye and help to identify
corresponding jet components. 
}
}
\end{minipage}
~~
\begin{minipage}[t]{6.5cm}{
\vspace{-16.5cm}
\psfig{figure=kri_fig2.epsi,width=6.0cm,angle=-90}
%\caption
{\footnotesize Fig.\ 2:
Spectral index variations along the jet. For 1997 (circles, solid lines),
the spectral index gradient is calculated directly from the intensity profiles of the
maps at 15 and 86\,GHz. For 1995 (squares, dashed line), the spectral indices were
derived from Gaussian component model fits at 22 and 86\,GHz. We note that 
during the time interval 1995 -- 1997, different jet components occupied this jet region.
The spectral profile along the jet, however, did not change significantly. 
}
\vspace{0.5cm}

\psfig{figure=kri_fig3.epsi,width=6cm,angle=-90}
%\caption
{\footnotesize Fig.\ 3:
The motion of the mean jet axis in the inner jet of 3C\,273 at 15\,GHz. Symbols denote
for different epochs: 1995.54 (circles), 1996.94 (squares), 1997.04 (diamonds),
and 1997.19 (triangles). The transverse oscillations of the ridge line are measured 
relative to a straight
line oriented along $\rm{pa}=240\deg$. We note the systematic longitudinal
displacement of maxima and minima. 
This corresponds to an apparent pattern velocity of $\beta_{app} \simeq 4.2$,
which is by a factor of up to 2 slower than the component motion.
}
}
\end{minipage}
\end{figure}

Dual-frequency maps obtained in 1995 (22/86\,GHz) and 1997 (15/86\,GHz) allow to measure
the spectral index gradient along the jet (see Fig.\ 2). The spectrum oscillates
between $-1.0 \leq \alpha \leq +0.5$ (\sprop).  Most noteworthy, the spectral gradients did not change
significantly over the 2 year time period, although different jet components occupied 
this jet region. Thus, the geometrical (eg. relativistic aberration) and/or the physical environment
(pressure, density, B-field) in the jet must determine the observed properties of the VLBI components.
Hence, the latter do not form `physical entities', but seem to react to the physical 
conditions of the jet fluid.

Further evidence for a fluid dynamical interpretation comes from a study
of the mean jet axis and the transverse width of the jet. Both oscillate
quasi-sinusoidally on mas-scales. At 15\,GHz, the variation of the ridge-line with time 
could be determined from 4 VLBA maps obtained during 1995 -- 1997 (Fig.~3). 
The maxima and minima of the ridge-line are systematically displaced. This `longitudinal'
shift suggests motion with a pattern velocity of $\beta_{app} =4.2$. 
The sinusoidal curvature of the jet axis, however, is more indicative for jet 
rotation rather than for longitudinal waves. 
Helical Kelvin-Helmholtz instabilities propagating in the jet sheath could mimic such 
rotation, which, when seen in projection, would explain 
also the spectral index oscillation in Figure 2.
\begin{figure}[t]
%\vspace{0.5cm}
\centerline{\psfig{figure=kri_fig4.epsi,width=10cm,angle=-90}}
%\caption
{\footnotesize Fig.\ 4:
Relative core separations r(t) for the components C6 -- C18.
The legend on the right identifies symbols with VLBI components. The lines
are least square fits to the data.
}
\vspace{-0.7cm}
\end{figure}

For many AGN, a correlation between flux density variability and ejection of VLBI components 
is suggested. In 3C\,273 we identified 13 jet components (C6 -- C18) and traced
their motion back to their ejection from the VLBI core. The typical measurement uncertainty
for the ejection times $t_0$ ranges between 0.2 -- 0.5\,yr. In Figure 5 (left) we plot $t_0$
and the millimeter-variability (22 -- 230\,GHz). We also add the Gamma-ray
detections of 3C\,273 from EGRET. 
In Figure 5 (right) we plot the onset times of the mm-flares derived
from these light curves (T\"urler et al. 1999) together with the VLBI ejection time ($t_0$) and the 
Gamma-ray fluxes. For each onset of a mm-flare, we find that a new jet component was ejected.

Although the time sampling of the Gamma-ray data is quite coarse,
a relation between component ejection and high Gamma-ray flux 
appears very likely (note that each Gamma-detection 
already means higher than usual $\gamma$-brightness). From a more detailed analysis
(Krichbaum et al. 2001) we obtain for the time lag between component ejection and onset of
a mm-flare: $t_0 - t_0^{\rm mm} = 0.1 \pm 0.2$\,yr. If we assume that the observed peaks in the Gamma-ray
light-curve are located near the times $t_0^\gamma$ of flux density maxima, we 
obtain  $t_0^\gamma -  t_0^{\rm mm} = 0.3 \pm 0.3$\,yr. Although the Gamma-ray variability
may be faster, this result is fully consistent with the more general finding of 
enhanced Gamma-ray fluxes mainly during the rising phase of millimeter flares.
We therefore suggest the following tentative sequence of events:
$t_0^{\rm mm} \leq  t_0 \leq t_0^{\gamma}$ -- the onset of a millimeter
flare is followed by the ejection of a new VLBI component and,
either simultaneously or slightly time-delayed, an increase of the Gamma-ray flux.
If we focus only on those VLBI components, which were ejected close to the main maxima of the 
Gamma-ray light-curve in Figure 5, we obtain time lags of $t_0^{\gamma} - t_0$
of $\leq 0.5$\,yr for C12, $\leq 0.9$\,yr for C13, $\leq 0.2$\,yr for
C16 and $\leq 0.1$\,yr for C18. In all cases the Gamma-rays seem to
peak a little later than the time of component ejection. With $\beta_{app} \simeq 4$
near the core, the Gamma-rays would then escape at a radius $r_\gamma \leq 0.1$\,mas.
This corresponds to $r_\gamma \leq 2000$ Schwarzschild radii (for a $10^9$ M\solar black hole) 
or $\leq 6 \cdot 10^{17}$\,cm, consistent with theoretical expectations, in which
Gamma-rays escape the horizon of photon-photon pair production at separations  
of a few hundred to a few thousand Schwarzschild radii.
\begin{figure}[h]
%\vspace{2cm}
\begin{minipage}[t]{6.5cm}{
\psfig{figure=kri_fig5a.epsi,width=6.5cm,angle=-90} 
}
\end{minipage}
~~
\begin{minipage}[t]{6.5cm}{
\psfig{figure=kri_fig5b.epsi,width=6.68cm,angle=-90}
}
\end{minipage}
%\caption
{\footnotesize Fig.\ 5:
Left: Flux density variations at 230 GHz (filled diamonds), 86 GHz (open circles)
and 22 GHz (filled squares). 
Upward oriented triangles denote Gamma-ray fluxes from EGRET,
downward oriented triangles are upper limits. The extrapolated
ejection times of the VLBI components and their uncertainties
are indicated by filled circles with horizontal bars along the time axis.
Right: Broad-band flux density activity and component ejection.
VLBI component ejection (open circles), Gamma-ray fluxes (triangles, 
downward oriented for upper limits) and
onset-times for the millimeter flares (open squares, 
from T\"urler et al. 1999). Labels denote the component identification from VLBI.
}
\end{figure}

\vspace{-3mm}
\noindent
{\bf References:}\\
{\small
T\"urler M., Courvoisier T.J.-L., \& Paltani S.,
`Modelling the submillimeter-to-radio flaring behaviour of 3C\,273', 1999, {\it A\&A}, {\bf 349}, p.~45--54.\\
Krichbaum T.P., et al., `High Frequency VLBI Observations of 3C\,273', 2001, {\it A\&A}, submitted.\\
}

\end{document}